\documentclass[structabstract, longauth, regular]{aa}  
\newcommand{\ttt}[1]{\times10^{#1}}

\renewcommand{\epsilon}{\varepsilon}

\usepackage{graphicx}
\usepackage{txfonts}
\usepackage{caption}
\usepackage{subcaption}

\usepackage{natbib}
\bibpunct{(}{)}{;}{a}{}{,} % to follow the A&A style

\begin{document}

   %\title{MAGIC observations of the Cataclysmic Variable AE Aqr during the 2012 multiwavelength campaign}
\title{MAGIC search for VHE $\gamma$-ray emission from AE Aquarii in a multiwavelength context}

% authors 24.04.2014  Format AA
%
\author{
J.~Aleksi\'c\inst{1} \and
S.~Ansoldi\inst{2} \and
L.~A.~Antonelli\inst{3} \and
P.~Antoranz\inst{4} \and
A.~Babic\inst{5} \and
P.~Bangale\inst{6} \and
J.~A.~Barrio\inst{7} \and
J.~Becerra Gonz\'alez\inst{8,}\inst{25} \and
W.~Bednarek\inst{9} \and
E.~Bernardini\inst{10}
% ...
}
\institute { IFAE, Campus UAB, E-08193 Bellaterra, Spain
\and Universit\`a di Udine, and INFN Trieste, I-33100 Udine, Italy
\and INAF National Institute for Astrophysics, I-00136 Rome, Italy
\and Universit\`a  di Siena, and INFN Pisa, I-53100 Siena, Italy
\and Croatian MAGIC Consortium, Rudjer Boskovic Institute, University of Rijeka and University of Split, HR-10000 Zagreb, Croatia
\and Max-Planck-Institut f\"ur Physik, D-80805 M\"unchen, Germany
\and Universidad Complutense, E-28040 Madrid, Spain
\and Inst. de Astrof\'isica de Canarias, E-38200 La Laguna, Tenerife, Spain
\and University of \L\'od\'z, PL-90236 Lodz, Poland
\and Deutsches Elektronen-Synchrotron (DESY), D-15738 Zeuthen, Germany
\and ETH Zurich, CH-8093 Zurich, Switzerland
\and Universit\"at W\"urzburg, D-97074 W\"urzburg, Germany
\and Centro de Investigaciones Energ\'eticas, Medioambientales y Tecnol\'ogicas, E-28040 Madrid, Spain
\and Institute of Space Sciences, E-08193 Barcelona, Spain
\and Technische Universit\"at Dortmund, D-44221 Dortmund, Germany
\and Universit\`a di Padova and INFN, I-35131 Padova, Italy
\and Unitat de F\'isica de les Radiacions, Departament de F\'isica, and CERES-IEEC, Universitat Aut\`onoma de Barcelona, E-08193 Bellaterra, Spain
\and Universitat de Barcelona, ICC, IEEC-UB, E-08028 Barcelona, Spain
\and Japanese MAGIC Consortium, Division of Physics and Astronomy, Kyoto University, Japan
\and Finnish MAGIC Consortium, Tuorla Observatory, University of Turku and Department of Physics, University of Oulu, Finland
\and Inst. for Nucl. Research and Nucl. Energy, BG-1784 Sofia, Bulgaria
\and Universit\`a di Pisa, and INFN Pisa, I-56126 Pisa, Italy
\and ICREA and Institute of Space Sciences, E-08193 Barcelona, Spain
\and Universit\`a dell'Insubria and INFN Milano Bicocca, Como, I-22100 Como, Italy
\and now at: NASA Goddard Space Flight Center, Greenbelt, MD 20771, USA and Department of Physics and Department of Astronomy, University of Maryland, College Park, MD 20742, USA
% ...
}

%\offprints{R.~L\'opez-Coto, \email{rlopez@ifae.es}, D.~Hadasch, \email{hadasch@ieec.uab.es}}

   %\date{Received September 15, 1996; accepted March 16, 1997}

\date{Received ... / Accepted ... Draft version \today}
%\date{}
% \abstract{}{}{}{}{} 
% 5 {} token are mandatory
 
  \abstract
   {It has been claimed that the nova-like cataclysmic variable (CV) AE Aquarii (AE Aqr) is a very-high-energy (VHE, $E>$100 GeV) source both on observational and theoretical grounds.}
   {We search for VHE $\gamma$-ray emission from AE Aqr during different states of the source at several wavelengths to confirm or rule out previous claims of detection of $\gamma$-ray emission from this object.
   }
   {We report on observations of AE Aqr performed by MAGIC. The source was observed during 12 hours as part of a multiwavelength campaign carried out between May and June 2012 covering the optical, X-ray, and $\gamma$-ray ranges. Besides MAGIC, the other facilities involved were the KVA, Skinakas, and Vidojevica telescopes in the optical and {\it Swift} in X-rays. We calculated integral upper limits coincident with different states of the source in the optical. We computed upper limits to the pulsed emission limiting the signal region to 30\% of the phaseogram and we also searched for pulsed emission at different frequencies applying the Rayleigh test}
   {AE Aqr was not detected at VHE energies during the multiwavelength campaign. We establish integral upper limits at the 95\% confidence level for the steady emission assuming the differential flux proportional to a power-law function $\mathrm{d} \phi/\mathrm{d} E \propto E^{-\Gamma}$, with a Crab-like photon spectral index of $\Gamma$=2.6. The upper limit above 200 GeV is 6.4$\times$10$^{-12}$ cm$^{-2}$s$^{-1}$ and above 1 TeV is 7.4$\times$10$^{-13}$  cm$^{-2}$s$^{-1}$. We obtained an upper limit for the pulsed emission of 2.6$\times$10$^{-12}$  cm$^{-2}$s$^{-1}$ for energies above 200 GeV. Applying the Rayleigh test for pulsed emission at different frequencies we did not find any significant signal. }
   {Our results indicate that AE Aqr is not a VHE $\gamma$-ray emitter at the level of emission previously claimed. We have established the most constraining upper limits for the VHE $\gamma$-ray emission of AE Aqr.}


   \authorrunning{Aleksi\'c et al.}
   \titlerunning{MAGIC search for VHE $\gamma$-ray emission from AE Aquarii in a multiwavelength context}
   \maketitle
%
%________________________________________________________________%

%\newpage

\section{Introduction}
\label{intro}

CVs are semi-detached binaries consisting of a white dwarf (WD) and a companion star (usually a red dwarf) that transfers matter to the WD. They are classified by the type of variation they manifest \citep[for a review see][]{CVs}. Since the discovery of transient $\gamma$-ray emission from the symbiotic nova V407 Cygni by \emph{Fermi}-LAT \citep{ScienceFermi} and the subsequent report of transient emission from four additional classical novae \citep{ArxivFermi,ATEL,ATELCen2013}, CVs have been included among high-energy emitters ($E>$100 MeV). 

AE Aqr is a bright nova-like cataclysmic binary consisting of a magnetic WD and a K4-5 V secondary. The orbital period of the system is $T_{\rm o}$=9.88 hours, and the spin period of the WD is $T_{\rm s}$=33.08 s, which is the shortest known for a WD \citep{Patterson79}. The system is located at a distance of 102$^{+42}_{-23}$ pc \citep{Distance}, and the spin-down power of the WD is 6$\ttt{33}$ erg s$^{-1}$  \citep{deJager94}.  It was originally classified as a DQ Her star \citep{Patterson94}, but it shows features that do not fit such a classification, e.g., violent variability at multiple wavelengths, Doppler tomograms that are not consistent with those of an accretion disk \citep{Doppler}, and the fast spin-down rate of the white dwarf  \citep[$\dot{P}$=5.64$\times10^{-14}$ s s$^{-1}$,][]{deJager94}. Recent X-ray measurements show that the spin-down rate is slightly higher, which is compatible with an additional term $\ddot{P}$=$3.46\ttt{-19}$ d$^{-1}$  \citep[][]{MaucheBreak}. AE Aqr is considered to be in a magnetic propeller phase, ejecting most of the material transferred from the secondary by the magnetic field of the WD \citep{Propeller}. It exhibits flares 50\% of the time, varying in the optical band from $B=12.5$~mag (during the low state) to $B=10$~mag (during flares). \cite{RelativisticElectrons} observed radio flares with fluxes in the range 1--12 mJy at 15 GHz. They show that the radio flares may be produced by relativistic electrons, which provides evidence of accelerated particles that radiate synchrotron emission in magnetized clouds. The time of the optical and radio flares is random. Soft (0.5--10 keV) and hard (10--30 keV) X-rays have also been detected with a 33 s-modulation \citep{X-ray_pulsation, MaucheBreak, SuzakuTerada}. A non-thermal origin of the hard X-rays is favored by \cite{SuzakuTerada}, who report an X-ray luminosity of  $L_{\text{Hard X-rays}}\simeq5\ttt{30}$ erg s$^{-1}$ for the isotropic emission. They also report a sharp feature in the hard X-ray pulse profile that has not been confirmed by subsequent observations \citep{Nustar}. Because of the large magnetic field and the fast rotating period of the WD, AE Aqr has been compared to pulsars \citep{Ikhsanov_pulsar} and has been proposed as a source of cosmic ray electrons \citep{Cosmic_electrons}. 

The groups operating the Nooitgedacht Mk I Cherenkov telescope \citep{Nooitgedacht_Cherenkov} and the University of Durham VilE gamma-ray telescopes \citep{Narrabri_Cherenkov} reported TeV $\gamma$-ray emission from AE Aqr using the imaging atmospheric Cherenkov technique. The Durham group claimed that they detected  $\gamma$-rays of energies above 350 GeV pulsed at the second harmonic of the optical period (60.46 mHz), as well as two bursts of TeV $\gamma$-rays \citep{Narrabri92,Chadwick_burst} lasting for 60 s and 4200 s with 4.5 $\sigma$ and 5.3 $\sigma$ significance, respectively. The Nooitgedacht group reported pulsed signals above energies of a few TeV at frequencies close to the spin frequency of the WD (30.23 mHz), with significances  varying from 3 $\sigma$ to 4 $\sigma$. \cite{Meintjes2012} claim that the duty cycle of the occurrence of TeV periodic signals above 95\% significance level is $\sim$ 30\%. They find coincidence in the orbital phase of their detections with the time of superior conjunction of the WD (orbital phase 0), but the burst reported by the Durham group is not coincident with this orbital phase. In the reports made by the two groups, the fluxes measured for the pulsed emission and burst episodes are at $10^{-9}$--$10^{-10}$ cm$^{-2}$s$^{-1}$ for $E>$350 GeV for the Durham group and $E>$2.4 TeV for the Nooitgedacht group. 

The luminosity corresponding to these fluxes is in the range $10^{32}$--$10^{34}$ erg s$^{-1}$, where the latter is at the level of the spin-down power of the WD. After the reports of TeV emission of such extraordinary luminosities, models were proposed to explain the fluxes measured \citep{PropellerModel}, as well as others predicting lower levels of emission \citep{EWD-model}. According to classical models of particle emission, the magnetic moment of some WDs in binaries might provide enough energy to accelerate particles to VHE \citep[][]{Acceleration}. The flux levels reported by the Durham and Nooitgedacht groups is measurable in less than one hour of observations with the current generation of Imaging Atmospheric Cherenkov Telescopes (IACTs). AE Aqr has been observed by different generations of IACTs since the detection claims were reported, but none have confirmed them. The Whipple telescope observed the source for 68.7 hours and did not find any evidence of emission \citep{AEAqr_Whipple}. They reported flux upper limits (U.L.) at 4$\times10^{-12}$ cm$^{-2}$s$^{-1}$ for the steady emission and 1.5$\times10^{-12}$ cm$^{-2}$s$^{-1}$ for the pulsed emission above 900 GeV. Later attempts by MAGIC and HESS did not lead to conclusive results \citep{Sidro,MaucheMWL}.
  
The purpose of this campaign was to obtain good results about the VHE emission of AE Aqr with MAGIC in a multiwavelength context, and hence confirm or rule out previous claims of $\gamma$-ray emission. We present in this work the results of the campaign, with emphasis on the search for signals in the VHE $\gamma$-ray range.

 %----------------------------------------------------------------------------------------------------------------%

 \section{Observations}
During the period between May 15 (MJD 56062) to June 19, 2012 (MJD 56097), we carried out a multiwavelength campaign to observe AE Aqr. The purpose of this campaign was to look for $\gamma$-ray emission during the different states of the source at several wavelengths. The log of the observation times during the campaign for all the instruments is shown in Table~\ref{Observation_times}. 

\begin{table*}[!t]
\begin{center}
\caption{Observation start and stop UT times for every night and every facility involved in the multiwavelength campaign. The number of minutes simultaneous to the MAGIC observations is included in brackets for each facility. }
\label{Observation_times}
\begin{tabular}{llllll}

\hline

\hline \hline
    Date [MJD]     & \multicolumn{1}{c}{KVA} & \multicolumn{1}{c}{Skinakas}  & \multicolumn{1}{c}{Vidojevica}   & \multicolumn{1}{c}{{\it Swift}} & \multicolumn{1}{c}{MAGIC}\\
\hline
   56062  & -  & - & - & 04:35 -- 04:54 & -\\
   \hline
   56063    & - &  - & - & 04:15 -- 04:49 &  -\\
   \hline
   56064   & 03:39 -- 05:07 & - & - & 03:04 -- 03:24 &  -\\
   \hline
   56065   & -  & - & - & 03:04 -- 03:22 & -  \\
      \hline
   56066  & -    & - & - & 03:10 -- 03:28 & -  \\
      \hline
   56067   & - & - & -  & 03:15 -- 03:33 &   - \\
   \hline
   56068  & 03:20 -- 04:18 &  - & - & 03:19 -- 03:37 & -  \\
   \hline
   56069    & 03:13 -- 04:14  & - & -  & 03:23 -- 03:41& -\\
   \hline
   56071  & -   & - & - & 03:30 -- 03:49 & -\\
\hline
  56072  & -   & - & - & 03:33 -- 03:54 & - \\
\hline  
   56073   &  02:50 -- 03:51 [43]  &  - & - & 03:37 -- 03:55& 02:47 -- 03:33 \\
\hline
   56074    &  02:47 -- 03:54 [38]&  - & - & 03:40 -- 03:57& 02:40 -- 03:25 \\
\hline
   56075   &  02:38 -- 03:39 [43]  &   - & -  &  02:11 -- 03:55 [43] & 02:40 -- 03:23 \\
\hline
   56076   &   02:48 -- 03:55 [29] &  - & - & 02:06 -- 02:26 & 02:34 -- 03:17  \\
\hline
   56077   &  02:51 -- 03:52 [23] &   - & - & 02:17 -- 02:36 [8]  &  02:28 -- 03:14\\
\hline
  56078   &  03:21 -- 04:22  &   - & - & 02:17 -- 02:36 [19] &  02:15 -- 03:09\\
\hline
   56079    &  03:24 -- 04:50 [73] &   - & - &  03:56 -- 04:15 [19] & 03:37 -- 04:57\\
\hline
  56080    &  03:57 -- 04:59 [57] & - & -& -  &  03:42 -- 04:54\\
\hline
   56090  &01:47 -- 02:30 & -  & - & -  & -  \\
\hline
   56091   & 01:49 -- 02:25  & - & -  & 01:28 -- 01:47 & -\\
\hline
   56092   &  01:33 -- 02:34 [61] & - & - &  01:13 -- 01:32& 01:32 -- 02:34\\
\hline
   56093    &  01:27 -- 02:16 [39]  &  01:17 -- 02:18 [39] &  01:13 -- 01:58 [27] & 01:15 -- 01:34 [3] &  01:31 -- 02:10\\
\hline
   56094   &  01:25 -- 02:07 [41]  &  01:03 -- 02:11 [43] &  00:21 -- 02:00 [37]  &  01:15 -- 01:35 [12] & 01:23 -- 02:06\\
\hline
   56095      &  01:20 -- 02:03 [38]  &  01:19 -- 02:18 [39] &   01:00 -- 02:00 [40] &  - & 01:18 -- 01:58\\
\hline
   56096  &  01:15 -- 01:54 [39]  &  01:11 -- 02:12 [51] &    00:35 -- 02:00 [47] & - & 01:13 -- 02:04 \\
\hline
   56097    &  01:11 -- 02:00 [33] & - &  01:02 -- 01:47 [25]  &  - & 01:22 -- 01:55\\
   \hline

\end{tabular}
\end{center}
\end {table*}

\subsection{Optical facilities}
We used data from three optical telescopes for the campaign. The observations are described in the following:

\subsubsection*{KVA}
The KVA optical telescope is located on La Palma, but is operated remotely from Finland. The telescope has a mirror diameter of 35 cm. The effective aperture ratio of the system is f/11 with a SBIG ST-8 CCD camera (0.98 arcsec/pix)  \citep{KVA}. 

The AE Aqr observations were performed in the $B$ band using 20-second exposures extending to about two hours of data per night during 19 nights. The magnitude of the source was measured from CCD images using differential photometry with 5" radius aperture, and the data were reduced using the standard analysis software to analyze KVA data \citep{KVA_software}. The seeing conditions during the observations were 1" FWHM. The typical error in the magnitude measurement is $\sim$0.04 mag. The comparison star used to calibrate the AE Aqr flux was the star 122 of the AAVSO AE Aqr finder chart.

\subsubsection*{Skinakas}
The data from the Skinakas Observatory in Crete (Greece) were obtained with the 1.3-m Ritchey-Chr\'etien telescope located on the Skinakas mountain at an altitude of 1750 meters. \footnote{http://skinakas.physics.uoc.gr/en/} The telescope has a focal ratio of f/7.6. The data were acquired with an Andor Tech DZ436 2048x2048 water cooled CCD. The physical pixel size is 13.5 microns resulting in 0.28 arcsec on the sky. The camera was used in the 2-$\mu$s-per-pixel readout mode. The observations were taken with a Bessel $B$ filter using 10-second exposures, while the cycle time from the start of one exposure to the next was 14 seconds.

The data from Skinakas were taken during about one hour for four nights, and they were reduced using IRAF routines. Differential photometry was performed using the photometry package DAOPHOT using 25 pixel (7") radius apertures. The seeing conditions during the observations were 2" FWHM. The typical error in the magnitude measurement is $\sim$0.005 mag. The AE Aqr data were flux-calibrated using stars 122 and 124 in the AAVSO AE Aqr finder chart.

\subsubsection*{Vidojevica}
The Astronomical Station Vidojevica is located on Mt. Vidojevica (Serbia), at an elevation of 1150 m. The data were obtained with the 60-cm Cassegrain telescope. \footnote{http://belissima.aob.rs/} The telescope was used in the f/10 configuration with the Apogee Alta U42 CCD camera (2048 x 2048 array, with 13.5-micron pixels providing a 0.46 arcsec/pix plate scale). The $B$ filter from Optec Inc. (Stock No. 17446) was used for all observations. The field centered on the target AE Aqr was observed continuously with ten seconds of exposure time. Only a fraction of the full CCD chip field-of-view, roughly 5 arcmin on a side, was read out in approximately four seconds, resulting in 14 seconds of total cycle time between exposures.

The data were taken for periods between one and two hours for five nights and they were reduced using standard procedures in IRAF. The photometry was performed with Source Extractor, using five-pixel (2.3") radius circular apertures. Typical seeing conditions during the observations were 2" FWHM. The typical error in the magnitude measurement is $\sim$0.015 mag. The AE Aqr flux was calibrated using the same comparison stars as for Skinakas.

\subsection{Swift}

{\it Swift} \citep{Swift} target-of-opportunity observations of AE Aqr were scheduled during 25 orbits from MJD 56062 to 56079 and from MJD 56091 to 56094. Data were obtained with the X-ray Telescope  \citep[XRT, sensitive over the energy range 0.3--10 keV][]{XRT}, the Ultraviolet/Optical Telescope (UVOT), and the Burst Alert Telescope (BAT), although only the XRT data have been analyzed to support the MAGIC observations. The screened and calibrated XRT CCD/PC event data for ObsIDs 00030295011--00030295035 were downloaded from the HEASARC data archive\footnote{http://heasarc.gsfc.nasa.gov/docs/archive.html}. The data was processed using a flexible IDL script developed by C. W. Mauche to deal with event data from instruments on numerous science satellites including {\it ROSAT\/}, {\it ASCA\/}, {\it EUVE\/}, {\it Chandra\/}, and {\it XMM-Newton\/}. The analysis was crosschecked using the XRTDAS software package (v.2.9.3) developed at the ASI Science
Data Center (ASDC) and distributed by HEASARC within the HEASoft package (v. 6.15.1). On-source events were selected within a circle of a 30-pixel (69 arcsec) radius. The background was evaluated in an adjacent 60-pixel radius off-source region. Event energies were restricted to the 0.5--10 keV bandpass, and all event times were corrected to the solar system barycenter. The 25 ObsIDs consisted of 29 good-time intervals, which were combined into 25 satellite orbits, although one orbit was rejected because the exposure was too short (20 s), and three orbits were rejected because the source image fell on one of the dead strips on the detector. The net exposure during the remaining orbits ranged from 559 s to 1178 s, with $\sim 950$ s being typical, and the total exposure was 19.94 ks.

\subsection{MAGIC}
\label{magic_intro}
MAGIC is an IACT situated on the Canary island of La Palma, Spain (28.8$^\circ$N, 17.9$^\circ$ W at 2225 m a.s.l). It is a stereoscopic system of two telescopes that achieves a sensitivity of (0.76 $\pm$ 0.03)\% of the Crab Nebula flux above 290 GeV in 50 hours \citep{Performance}. Its energy threshold for observations at low zenith angles is 50 GeV.

MAGIC observed AE Aqr during 14 non-consecutive nights during the period between MJD 56073 and 56097. The observations were performed with a single telescope owing to a hardware failure in MAGIC I camera. This worsened the sensitivity to $\sim$1.5\% of the Crab Nebula flux above 300 GeV in 50 hours \citep{Performance}. The source was observed at zenith angles ranging between 28$^\circ$ and 50$^\circ$, and after quality cuts, 9.5 hours of data were obtained. The data were taken in \emph{wobble mode} pointing at two different symmetric regions situated 0.4$^\circ$ away from the source to evaluate the background simultaneously with AE Aqr observations \citep{Wobble}. They were analyzed using the MARS analysis framework \citep{MARS}. The gamma/hadron separation, the event direction reconstruction, and the energy estimation of the primary gamma event were done using the random forest method \citep{RF}. To calculate flux U.L. for steady emission, we used the Rolke algorithm \citep{Rolke} with a confidence level (C.L.) of 95\% assuming a Gaussian background and 30\% systematic uncertainty in the efficiency of the $\gamma$-ray selection cuts.

To search for pulsed emission, the arrival times of the events were corrected to the solar system barycenter using the software package TEMPO2 \citep{TEMPO2}. To calculate the phases of the events, we used the ephemeris presented in  \cite{deJager94} using the second-order correction proposed by \cite{MaucheBreak}. We corrected the times for the orbital motion of the system using TEMPO2 as well. The ephemeris, particularly the phase of spin-pulse maximum, was checked using the Swift data (see Sect. \ref{x-rays}). The U.L. for the pulsed emission were calculated with a 95\% C.L. following the method described in \cite{deJager_pulsed_ULs} that makes use of the H-test for the significance of weak periodic signals \citep{H-test}. The simultaneity of the optical and MAGIC observations allows us to investigate the TeV flux of the source at different optical emission levels.

 %--------------------------------------------------------------------------------------------------------------%
 
 \section{Results}

\begin{figure}[]
 \centering
 \includegraphics[width=0.5\textwidth]{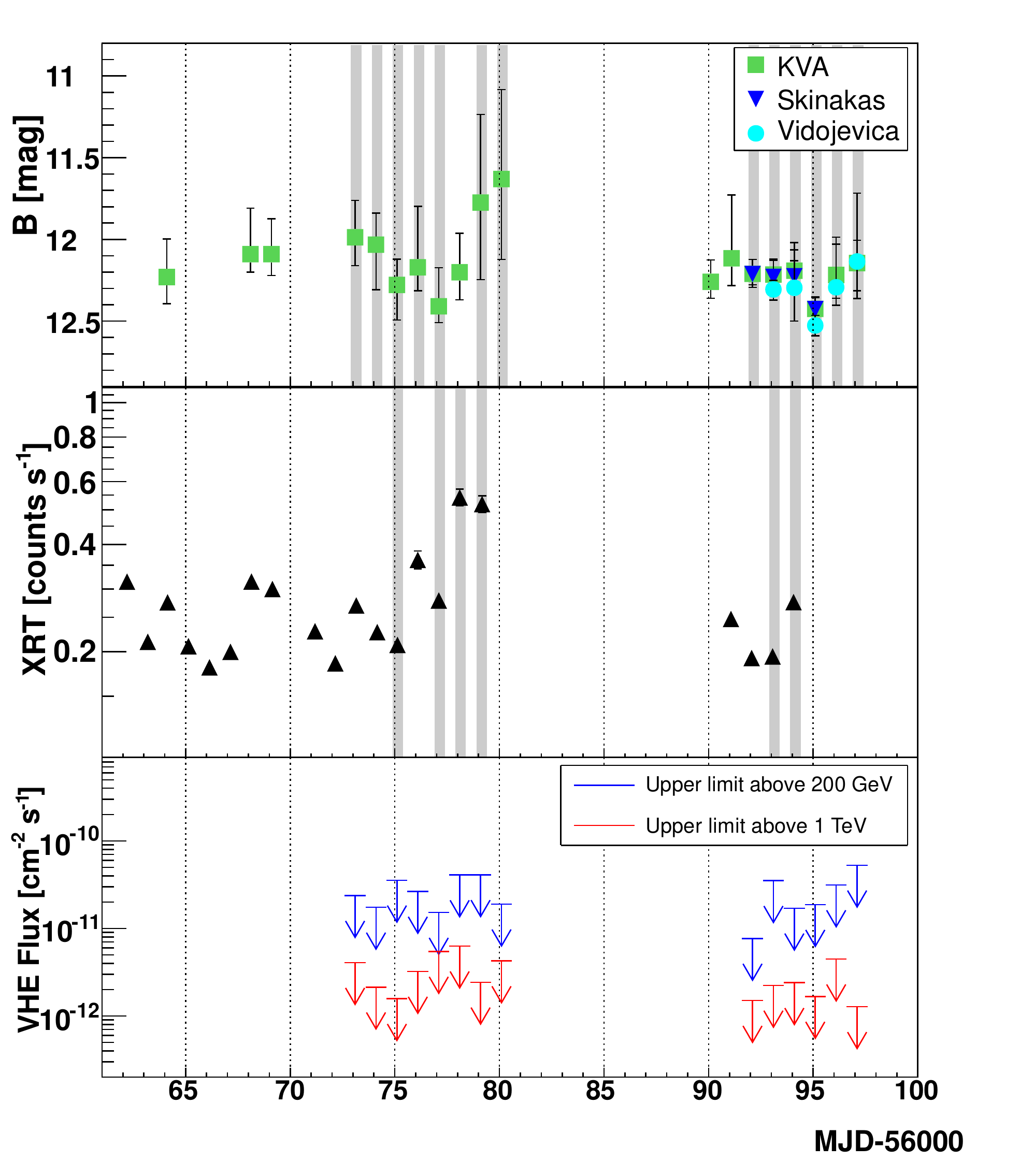}
\caption{Light curves of the multiwavelength campaign. The plot includes $B$ magnitudes measured by the optical telescopes (top), XRT count rate in the energy range 0.5--10 keV (middle) and MAGIC daily integral U.L. assuming a power-law spectrum with a 2.6 photon spectral index above 200 GeV and 1 TeV (bottom). Vertical dotted lines every 5 days are plotted across all the panels for reference. For the optical data, since the source variability is very large, the point plotted is the average magnitude of the night and the error bars indicate the maximum and minimum magnitude reached during that observation night. The shaded areas indicate the X-ray and optical observations with simultaneous data with MAGIC.}
  \label{LC}
 \end{figure}

The measured optical magnitudes are presented in section \ref{optical}. In section \ref{x-rays} the results obtained with Swift are discussed. In section \ref{MAGIC}, we present the results of the search for a steady and pulsed $\gamma$-ray signal. A summary of the observation logs of all the facilities can be found in Table~\ref{Observation_times}. The light curves of the multiwavelength campaign are shown in Fig.~\ref{LC}.

\subsection{Optical results}
\label{optical}
We present the results of all the optical observations together to check for consistency between the magnitudes measured by the different telescopes (upper panel of Fig.~\ref{LC}). The highest optical state was measured on MJD 56080, reaching $B=11.08$~mag. The short time exposures ($\sim$ 10 seconds) mean that it is not possible to produce the optical spin-phase-folded light curve.

 \subsection{Swift results}
\label{x-rays}

\begin{figure}[!t]
  \centering
  \includegraphics[width=0.5\textwidth]{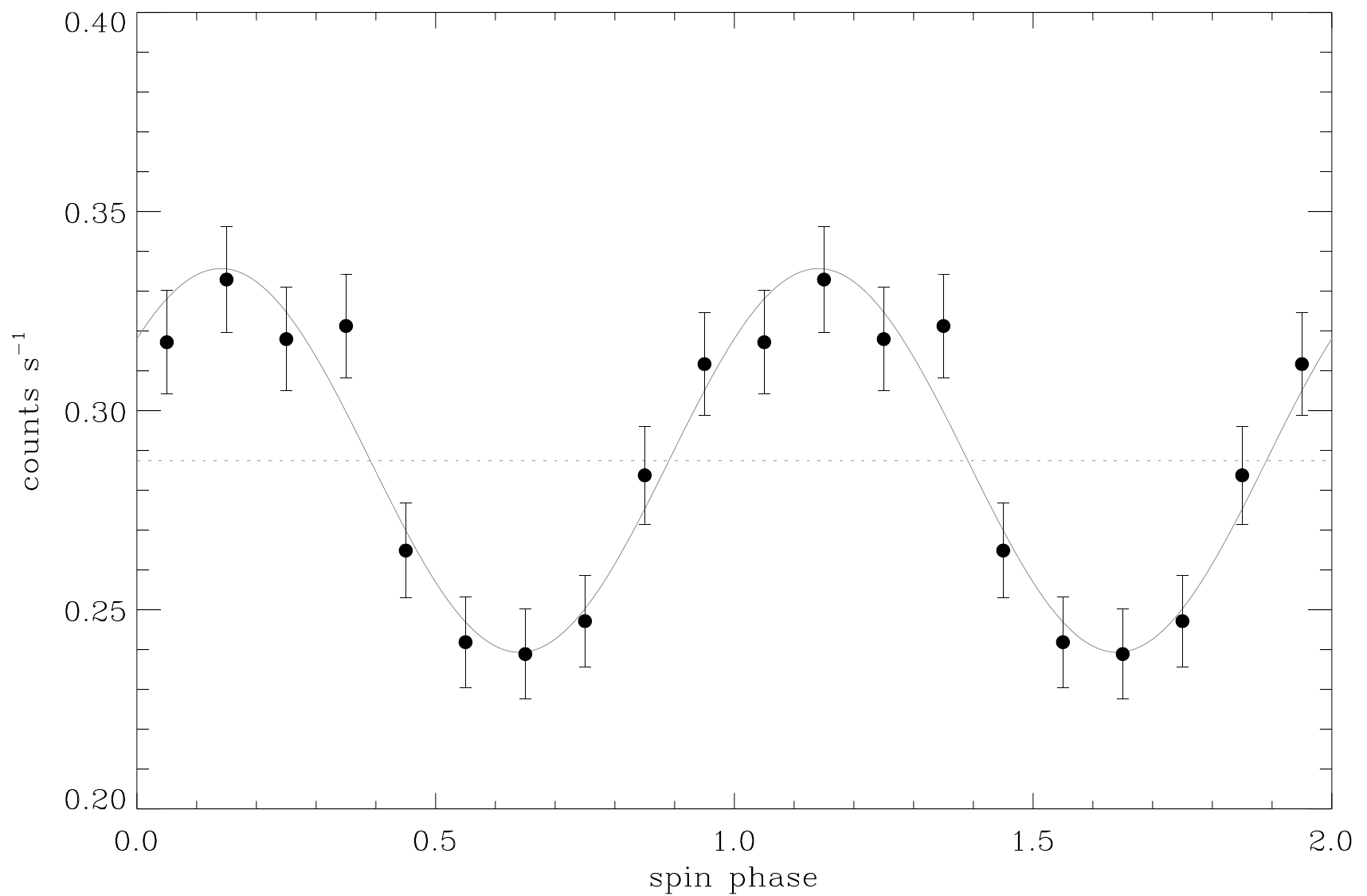}
  \caption{XRT spin-phase-folded light curve in the energy range 0.5--10 keV. Two cycles are shown for clarity. The errors quoted are the square root of number of counts in the source region plus the area-scaled number of counts in the background region, divided by the exposure. The continuous black line shows the best fit with a cosine $A(\phi_{\text{spin}}$) function. The dashed line represents the mean value $A_0$=0.260 counts s$^{-1}$ obtained from the fit.}
  \label{XRT_spin}
 \end{figure}

The {\it Swift\/}/XRT event data were used to compute the X-ray light curve (Fig.~\ref{LC}, middle) and the spin-phase-folded light curve  (Fig.~\ref{XRT_spin}). The background-subtracted XRT count rate varied by a factor of three, from 0.18 counts~s$^{-1}$ to 0.53 counts~s$^{-1}$, with a mean of 0.27 counts~s$^{-1}$. A similar ratio of mean-to-base and peak-to-base count rate ratios and a similar light curve morphology were observed during the long {\it Chandra\/} observation of AE Aqr in 2005  \citep{Mauche_2009}. The spin-phase-folded light curve was calculated using the ephemeris provided by  \cite{MaucheBreak}, with parameters:

$$
 \begin{array}{ll}
  \text{Orbital period } & P_{\text{orb}}  = 0.411655610 \text{ d}\\
  \text{Time of superior conjunction } & T_{0}  = 2445172.2784 \text{ BJD}\\
  \text{Spin period } & P_{\text{s}}  = 0.00038283263840 \text{ d}\\
  \text{Spin period derivative } &\dot{P}_{\text{s}}  =5.642\ttt{-14} \text{ d } \text{d}^{-1}\\
    \text{Spin period second derivative } & \ddot{P}_{\text{s}} = 3.46\ttt{-19} \text{ d}^{-1}\\
   \text{Projected semi-amplitude } & a_{\text{WD}}\sin i = 2.04 \text{ s.}\\

\end{array}
 $$

The points were fit with a cosine function $A(\phi_{\text{spin}}$)=$A_0$+$A_1$cos[2$\pi$($\phi_{\text{spin}}$-$\phi_{\text{off}}$)] with

\begin{eqnarray*}
 A_{0}& =& 0.260\pm 0.004 \text{ counts s}^{-1}\\
 A_{1} &= &0.042\pm 0.005 \text{ counts s}^{-1}\\
 \phi_{\text{off}} &=& 0.15\pm 0.02
\end{eqnarray*}

\noindent and $\chi^2$/dof=5.90/7=0.84. The fit function is shown in Fig.~\ref{XRT_spin}.

As a result, the relative pulse amplitude is $A_1$/$A_0$=16\%$\pm$2\%, which is slightly higher than previously measured by {\it ASCA}, {\it XMM-Newton}, and { \it Chandra}, which are 13\%, 10\%, and 15\%, respectively \citep[see Table~2 of][]{MaucheBreak}. A shift of $\phi_{\mathrm{off}}$=0.15$\pm$0.02, which is not compatible with $\phi_{\mathrm{off}}$=0, is observed. That is an indication of the inaccuracy of the ephemeris used or a drastic variation in either $\dot{P}$ or $\ddot{P}$. Nevertheless, we use this result for the time of the maximum of the pulsed X-ray emission to look for pulsed gamma-ray signals.

\subsection{MAGIC results}
\label{MAGIC}
The peak of the true energy distribution of gammas simulated with a power law with a 2.6 photon spectral index is 250 GeV, although the number of events surviving the g/h separation cuts is still high below this energy, down to 200 GeV, where it falls rapidly. This result is obtained from Monte Carlo simulations without applying any cut in reconstructed energy. We searched for steady and periodic emission in the MAGIC dataset. We computed U.L. to the integral flux above two values of energy; namely, above 200 GeV, as the lowest energy with sufficient gamma-ray detection efficiency (for this observation); and above 1 TeV, to compare our results with the previous claims. Most of those observations were simultaneous with optical and X-ray ones. Therefore, we also study the correlation of optical/X-ray flux with the possible $\gamma$-ray emission.

\subsubsection{Search for steady TeV emission}
\label{Steady}

 The total dataset did not show any significant steady signal. For the U.L. calculation, we assumed power-law functions with different photon spectral indices (2.0, 2.6, and 3.0).  The results are listed in Table~\ref{Slopes}. We also computed integral U.L. (95\% C.L.) for the single-night observations, assuming a source steady emission with a power-law function with a photon spectral index of 2.6. These U.L. can be found in Table~\ref{daily_ULs} and are plotted in Fig.~\ref{LC} (bottom panel). The single-night U.L. for TeV emission coincident with the highest states of the source in X-rays (MJD 56078 and 56079) and in the optical (MJD 56079 and 56080) are at the same level as the U.L. for the remaining days.

\begin{table}[!t]
\begin{center}
\caption{MAGIC integral U.L. to steady flux assuming a power-law spectrum with different photon spectral indices $\Gamma$ above 200 GeV and 1 TeV.}\label{Slopes}

\begin{tabular}{c c c }

\hline \hline
  &  \multicolumn{2}{c}{U.L. (95 \% C.L.)}\\
  $\Gamma$  & \multicolumn{2}{c}{$[$cm$^{-2}$s$^{-1}]$}    \\
   \cline{2-3}
 & $>$ 200 GeV & $>$ 1 TeV \\
\hline    
 2.0   & 4.2$\times$10$^{-12}$& 7.6$\times$10$^{-13}$ \\
 2.6    & 6.4$\times$10$^{-12}$  & 7.4$\times$10$^{-13}$\\
 3.0   & 8.0$\times$10$^{-12}$ & 7.4$\times$10$^{-13}$  \\
\hline
\end{tabular}
\end{center}
\end {table}

\begin{table}[!t]
\begin{center}
\caption{MAGIC daily integral U.L. to steady flux assuming a power-law spectrum with a photon spectral index of 2.6 above 200 GeV and 1 TeV.}\label{daily_ULs}
\begin{tabular}{ccccc}

\hline
\hline
  Date [MJD] &  \multicolumn{2}{c}{U.L. (95 \% C.L.)}\\
   & \multicolumn{2}{c}{$[$cm$^{-2}$s$^{-1}]$}    \\
\hline
 & $>$ 200 GeV & $>$ 1 TeV \\

\hline

56073 & 2.4$\ttt{-11}$& 4.0$\ttt{-12}$ \\
56074 &1.7$\ttt{-11}$ & 2.1$\ttt{-12}$ \\
56075 & 3.6$\ttt{-11}$ & 1.6$\ttt{-12}$\\
56076 & 2.7$\ttt{-11}$ & 3.2$\ttt{-12}$ \\
56077 & 1.5$\ttt{-11}$& 5.5$\ttt{-12}$ \\
56078 & 4.1$\ttt{-11}$ & 6.3$\ttt{-12}$\\
56079 & 4.1$\ttt{-11}$ & 2.4$\ttt{-12}$ \\
56080 & 1.9$\ttt{-11}$ & 4.3$\ttt{-12}$ \\
56092 & 0.8$\ttt{-11}$ & 1.5$\ttt{-12}$ \\
56093 & 3.5$\ttt{-11}$ & 2.2$\ttt{-12}$ \\
56094 & 1.7$\ttt{-11}$ & 2.4$\ttt{-12}$\\
56095 & 1.9$\ttt{-11}$ & 1.7$\ttt{-12}$ \\
56096  & 3.1$\ttt{-11}$ & 4.5$\ttt{-12}$ \\
56097 & 5.3$\ttt{-11}$ &1.3$\ttt{-12}$ \\

\hline
\end{tabular}
\end{center}
\end {table}

We also studied the behavior of the source during different bright optical states. Based on the optical states observed during the multiwavelength campaign, we selected $\gamma$-ray events during times when $B<12$~mag (1.22 hours) and $B<11.5$~mag (0.34 hours). The integral U.L. for those states are shown in Table~\ref{UL_optical}.

\begin{table}[!t]
\begin{center}
\caption{MAGIC integral U.L. to steady flux for different optical states above 200 GeV and 1 TeV and for a photon spectral index 2.6.}\label{UL_optical}
\begin{tabular}{ccc}

\hline \hline
   &  \multicolumn{2}{c}{U.L. (95 \% C.L.)}\\
 $B$ [mag] & \multicolumn{2}{c}{$[$cm$^{-2}$s$^{-1}]$}    \\
 \cline{2-3} 
 & $>$ 200 GeV & $>$ 1 TeV\\
\hline    
$<$ 11.5   & 2.1$\times$10$^{-11}$ & 1.6$\times$10$^{-12}$ \\
$<$ 12    & 7.3$\times$10$^{-12}$ & 1.2$\times$10$^{-12}$\\
\hline
\end{tabular}
\end{center}
\end {table}

\subsubsection{Search for pulsed TeV emission}
\label{pulsed_section}
\begin{figure}
        \centering
        \begin{subfigure}[b]{0.5\textwidth}
                \centering
             \includegraphics[width=\textwidth]{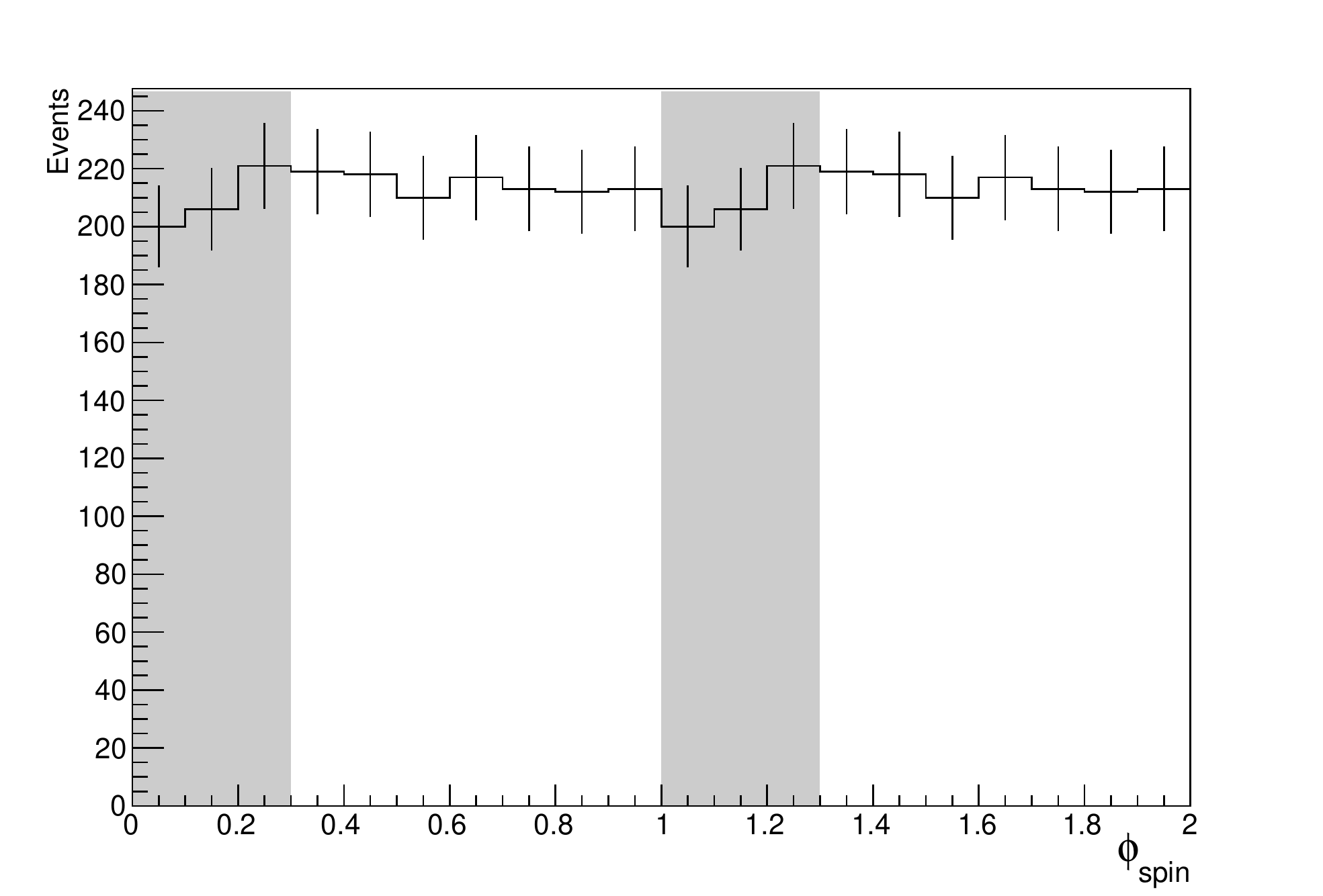}
        \end{subfigure}%

        \begin{subfigure}[b]{0.5\textwidth}
                \centering
                  \includegraphics[width=\textwidth]{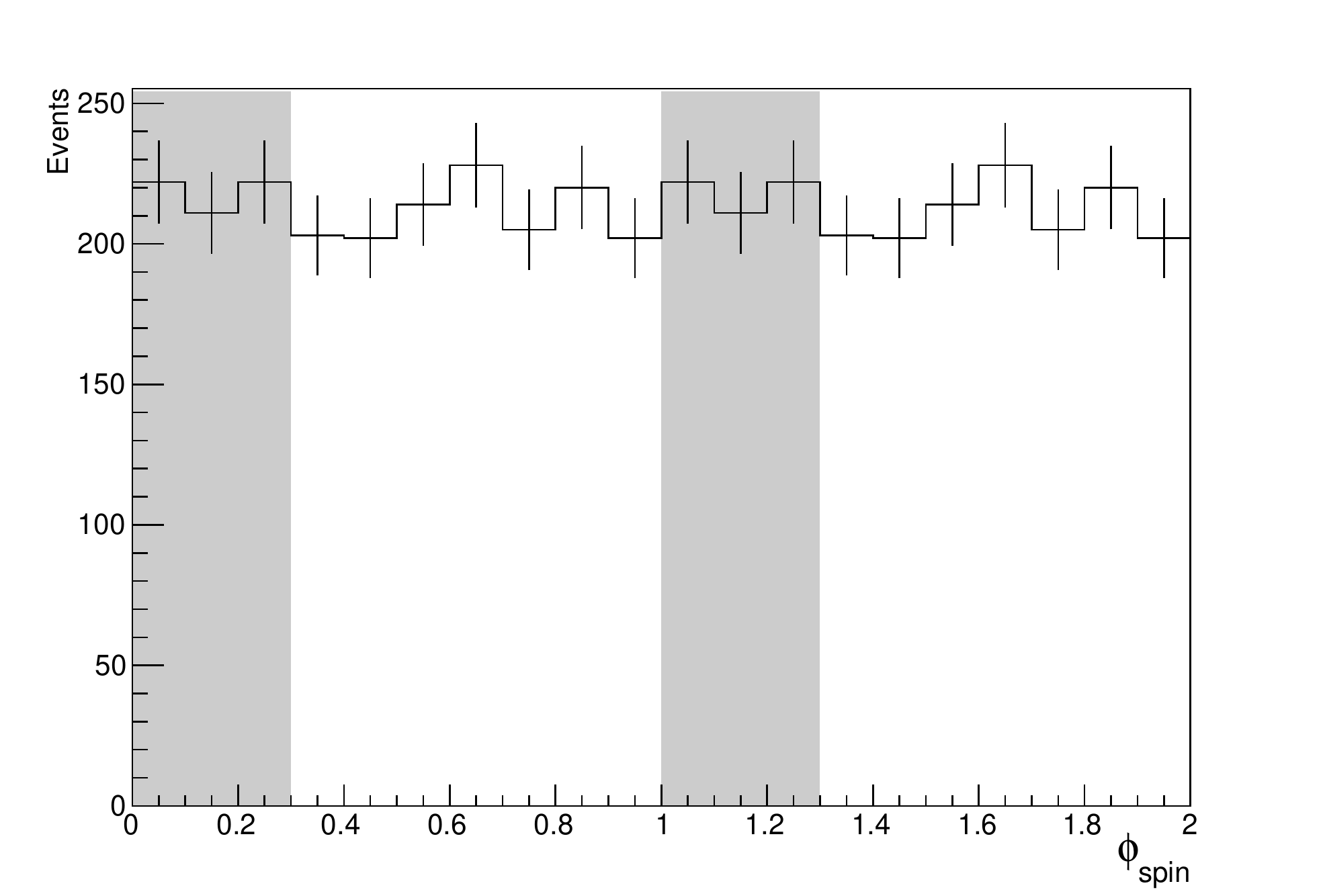}
        \end{subfigure}
        \caption{Phaseogram for the MAGIC data above 200 GeV for a frequency of 30.23 mHz (top) and 60.46 mHz (bottom). The shaded area corresponds to the region where the signal is expected assuming a duty cycle of 30\%.}\label{Phaseogram}
\end{figure}

We searched for pulsed TeV emission at the rotation frequency of the WD (30.23 mHz) and its first harmonic (60.46 mHz). We did not find any hint of periodic signal for any of the two frequencies.  For the upper-limit calculation, we limited the signal region to 30\% of the pulsar phaseogram, centered on the bin corresponding to the maximum of the XRT spin-phase-folded light curve (see Fig.~\ref{XRT_spin}). The phaseograms for data above 200 GeV are shown in Fig. \ref{Phaseogram}. These U.L., calculated as explained in section \ref{magic_intro}, can be found in Table~\ref{ULs_pulsed_emission}. 

\begin{table}[!t]
\begin{center}
\caption{MAGIC integral U.L. for the pulsed emission at the spin frequency and its first harmonic above 200 GeV and 1 TeV for a photon spectral index of 2.6.}\label{ULs_pulsed_emission}
\begin{tabular}{ccc}

\hline \hline
  &  \multicolumn{2}{c}{U.L. (95 \% C.L.)}\\
 Frequency  & \multicolumn{2}{c}{$[$cm$^{-2}$s$^{-1}]$}    \\
 \cline{2-3} 
 & $>$ 200 GeV & $>$ 1 TeV\\
\hline    
30.23 mHz   & 2.6$\times$10$^{-12}$ & 2.6$\times$10$^{-12}$ \\
60.46 mHz    & 2.1$\times$10$^{-12}$ & 3.7$\times$10$^{-12}$\\
\hline
\end{tabular}
\end{center}
\end {table}

\begin{figure}[]
 \centering
 \includegraphics[width=0.5\textwidth]{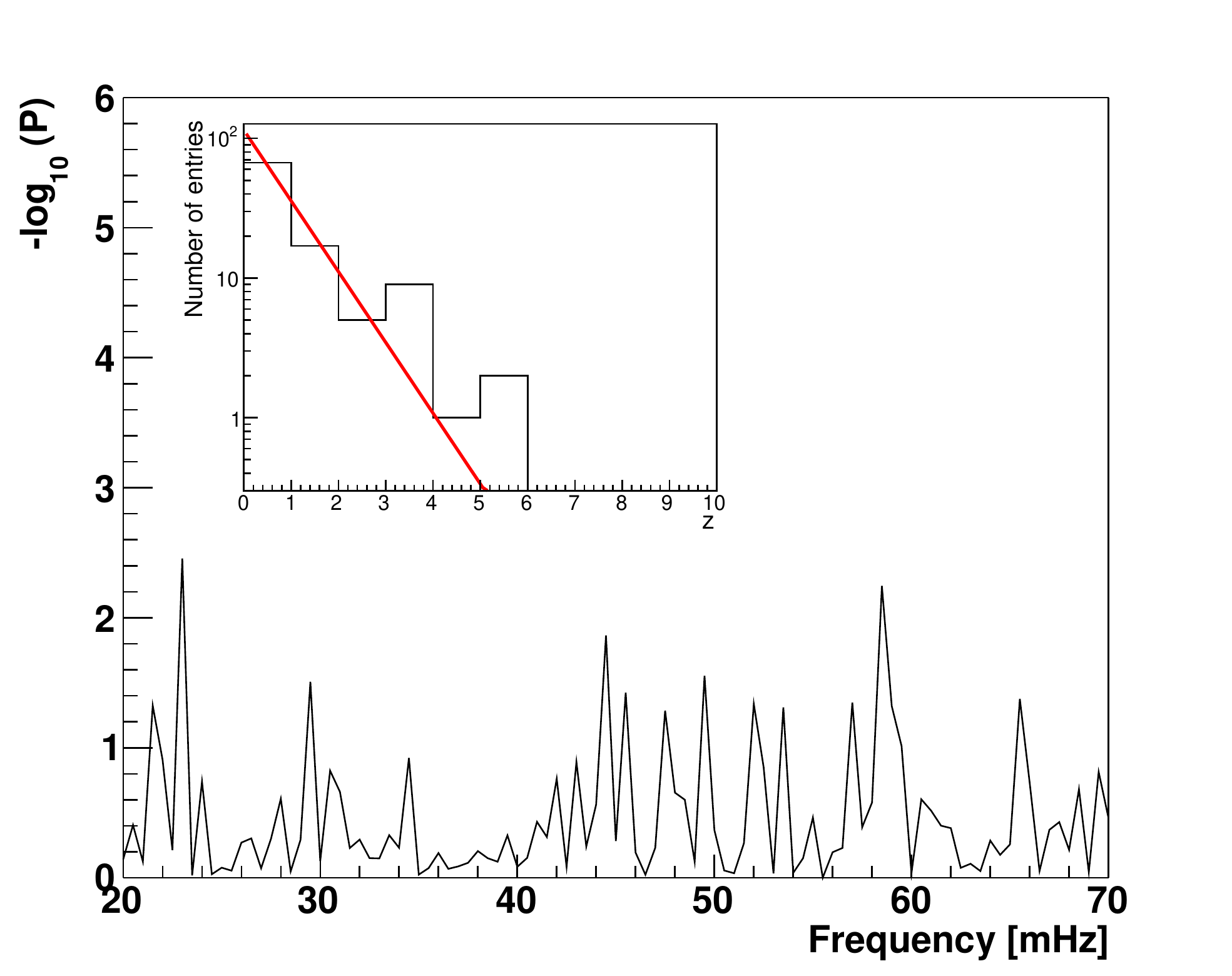}
 \caption{Periodogram of the frequencies in the range between 20.0 mHz and 70.0 mHz in steps of 0.5 mHz for the complete MAGIC dataset. The selected events have energies above 200 GeV. The plot in the inset represents the histogram of the Rayleigh power z for the complete MAGIC dataset.}
  \label{Rayleigh_plot}
 \end{figure}

We also searched for periodic emission at different frequencies using the Rayleigh test \citep{Rayleigh_test}. We scanned the complete dataset for periodic signals in the range between 20.0 mHz and 70.0 mHz in steps of 0.5 mHz (101 frequencies). This range is selected in order to cover the whole range of interest in the frequencies. For all the frequencies, we calculated the Rayleigh power $z$ and the chance probability of getting that value or higher from pure white noise as $P$ = $\exp$(-$z$). The histogram of $z$ values is fit with an exponential function $f(z)=A \exp(-bz)$. In case of purely white noise, we expect $b$=1 and $A=b \times N$, where $N$ is the number of scanned frequencies. The result of the fit is $A$=115$\pm$22 and $b$=1.17$\pm$0.17. The complete dataset scan for significant periodic signals is shown in Fig.~\ref{Rayleigh_plot}. The result of the fit of the histogram in the inset of Fig.~\ref{Rayleigh_plot} is compatible with white noise. The minimum pretrial chance probability obtained is 3.5$\ttt{-3}$ for a frequency of 23.0 mHz, which corrected after trials (101 frequencies) gives a post-trial probability of 3.0 $\ttt{-1}$. No significant signal of periodic/variable behavior was found.

\begin{figure}[]
 \centering
 \includegraphics[width=0.5\textwidth]{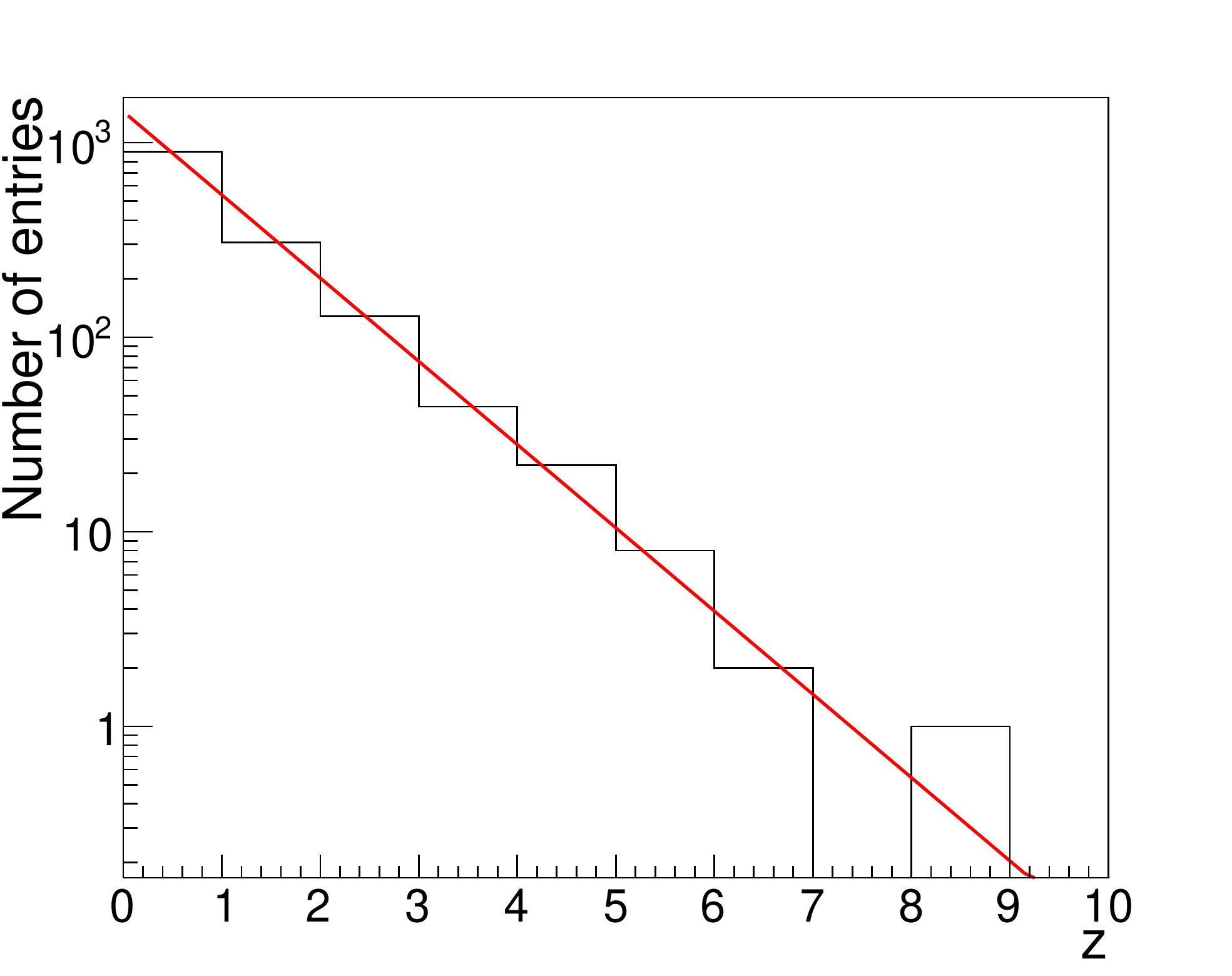}
 \caption{Histogram of the Rayleigh power $z$ for events above 200 GeV for all frequencies and individual MAGIC observations.}
  \label{Rayleigh_histogram}
 \end{figure}

We applied the Rayleigh test to the daily datasets as well. The range of frequencies is the same as the one used for the complete dataset. The minimum pretrial chance probability obtained for all the scans is 1.5$\ttt{-4}$ for a frequency of 54.0 mHz, achieved on MJD 56094. This probability, corrected after trials (101 frequencies $\times$ 14 observations), gives a 1.9$\ttt{-1}$ post-trial chance probability. The histogram with the distribution of Rayleigh power for all scanned frequencies and days is shown in Fig.~\ref{Rayleigh_histogram}. The result of the fit of the histogram $f(z)$ is $A$=1450$\pm$60 and $b$=0.99$\pm$0.03, which is compatible with white noise.

%__________________________________________________________________%

\section{Discussion}

Our MAGIC observations did not confirm the previous reports of emission from AE Aqr. We report flux values from two to three orders of magnitude below the fluxes previously reported. Specifically, \cite{Meintjes2012}, using the Nooitgedacht telescope, reported the detection of periodic signals with 95\% CL significance in 30\% of the observation time, and assuming that this is the typical behavior of the source, MAGIC should have observed similar fluxes in 30\% of the observation time. Thanks to the higher sensitivity of MAGIC with respect to the Nooitgedacht telescopes, those observations should have produced signals with much greater significance than the 95\% C.L. According to the results, we do not find any hint of a pulsed signal. Regarding the reports of random VHE bursts from the Narrabri telescope \citep{Narrabri92,Chadwick_burst}, since they do not follow any periodicity, they cannot be excluded by the results presented in this paper.

There are several days when the source is in a higher state in the optical and X-rays than the baseline. In the $B$ band, the source reached a state up to 1.5 magnitudes higher than the quiescence state at magnitude 12.5. In X-rays, the highest state is about three times higher than the baseline. This eruptive behavior is normal for this source. We searched for a correlation between the optical/X-ray emission and the $\gamma$-ray U.L. As shown in section \ref{Steady}, flux U.L. for the $\gamma$-ray emission are at the same level for all days, independently of the state of the source in the optical or X-rays. 

If we take the U.L. on the integral flux of Tables~\ref{Slopes} and \ref{ULs_pulsed_emission}, we can calculate the U.L. of the $\gamma$-ray luminosity of AE Aqr. The U.L. on the luminosity for the steady emission of AE Aqr, considering a power-law function with photon spectral index 2.6, above 200 GeV is $L_{\gamma, E>200 \text{GeV}}<$ 6.8 $\ttt{30}$ $d^2_{100}$ erg~s$^{-1}$, where $d^2_{100}$ is the distance normalized to 100 pc and above 1 TeV is $L_{\gamma, E>1 \text{TeV}}<$ 3.9 $\ttt{30}$ $d^2_{100}$ erg~s$^{-1}$. The U.L. on the luminosity for the pulsed emission at 30.23 mHz and 60.46 mHz, considering a power law with photon spectral index 2.6, above 200 GeV, are $L_{\gamma, E>200 \text{GeV}}$[30.23 mHz] $<$ 2.8 $\ttt{30}$ $d^2_{100}$ erg~s$^{-1}$  and $L_{\gamma, E>200 \text{GeV}}$ [60.46 mHz] $<$ 2.2 $\ttt{30}$ $d^2_{100}$ erg~s$^{-1}$. 

To explain the large $\gamma$-ray fluxes measured in the past, \cite{PropellerModel} proposed a model based on the propeller emission of particles that predicts large $\gamma$-ray fluxes, which are easily detected with the current generation of IACTs. The generation of VHE particles is based on the idea that a very high potential difference can be generated thanks to differences in the density of the gas present in a clumpy ring surrounding the WD. This model predicts luminosities of up to $L_{\gamma}\sim$10$^{34}$erg s$^{-1}$ during the largest bursts of the source, which would be able to explain the fluxes observed at $F$ $>$10$^{-10}$cm$^{-2}$ s$^{-1}$ by \cite{Meintjes94} and \cite{Chadwick_burst}. To explain these luminosities, the model makes assumptions that do not match the observations, like the presence of an accretion disk. We present in this paper U.L. on the pulsed/steady $\gamma$-ray luminosities measured by MAGIC on the order of $10^{30}$ erg s$^{-1}$, which is several orders of magnitude below the prediction of the model. 

Since there is evidence of non-thermal emission in the system, there has to be a mechanism that converts a fraction of the spin-down power into particle acceleration. To explain this non-thermal emission, there are mechanisms like the magnetic pumping in the magnetosphere \citep{Kuijpers}, which explains the radio outbursts as eruptions of bubbles of fast particles from the magnetosphere surrounding the WD, and the ejector white dwarf (EWD) model \citep{Ikhsanov_pulsar}, which describes a pulsar-like acceleration mechanism for AE Aqr and predicts the $\gamma$-ray emission of the system as well \citep{EWD-model}. Following the EWD model, the source emits TeV $\gamma$-rays during the optical highest state of the source ($B=10$~mag) with a luminosity lower than 4$\ttt{29}$erg~s$^{-1}$. The U.L. for higher optical magnitudes derived in this paper are one order of magnitude higher, therefore they do not conflict with our results. The future Cherenkov Telescope Array  \citep[CTA;][]{CTA} will have a sensitivity that is roughly one order of magnitude better than the current sensitivity of MAGIC \citep{PosterMC}. The flux prediction for the high-level optical state is expected to be detectable by CTA.

%__________________________________________________________________%

\section{Conclusions}

We carried out VHE observations of AE Aqr that were simultaneous to optical and X-ray ones, which allowed us to characterize the behavior of the source in different states. During our observations, the source displayed a level of brightness and type of variability that was consistent with previous observations in the optical and X-rays wavebands. We found a shift in the maximum of the spin-phase-folded X-ray light curve. This shift was unexpected according to the ephemeris used. We searched for steady $\gamma$-ray emission during the whole observation period, coincident with different optical states and also pulsed  $\gamma$-ray emission. We did not find any significant $\gamma$-ray emission from AE Aqr in any of the searches performed. We have established the most restrictive U.L. so far for the VHE emission (above 200 GeV and above 1 TeV) of this source, and of any other CV in general. The corresponding U.L. are up to three orders of magnitude lower than some of the emission reports by the Nooitgedacht and Durham groups about two decades ago. 
The propeller model is a good candidate for explaining the emission from radio to X-ray energies. However, it is very unlikely to be responsible for the production of $\gamma$-ray photons in the way described in \cite{PropellerModel}, unless the probability of flaring events is less than reported. Finally, we note that the level of $\gamma$-ray emission predicted by the EWD model is consistent with our U.L., and it could be detected with CTA.

%__________________________________________________________________%

\begin{acknowledgements}

We would like to thank
the Instituto de Astrof\'{\i}sica de Canarias
for the excellent working conditions
at the Observatorio del Roque de los Muchachos in La Palma.
The support of the German BMBF and MPG,
the Italian INFN, 
the Swiss National Fund SNF,
and the Spanish MINECO
is gratefully acknowledged.
This work was also supported
by the CPAN CSD2007-00042 and MultiDark CSD2009-00064 projects of the Spanish Consolider-Ingenio 2010 program,
by grant 127740 of the Academy of Finland,
by the DFG Cluster of Excellence ``Origin and Structure of the Universe'',
by the Croatian Science Foundation (HrZZ) Project 09/176,
by the DFG Collaborative Research Centers SFB823/C4 and SFB876/C3,
and by the Polish MNiSzW grant 745/N-HESS-MAGIC/2010/0.
CWM's contribution to this work was performed under the auspices of the U.S. Department of Energy by Lawrence Livermore National Laboratory under Contract DE-AC52-07NA27344. MB acknowledges support from the Serbian MESTD through grant ON176021. The authors thank N. Gehrels for approving our request for target-of-opportunity observations and the Swift Science Operations Team for scheduling them. This research made use of data provided by the HEASARC, which is a service of the Astrophysics Science Division at NASA/GSFC and the High Energy Astrophysics Division of the Smithsonian Astrophysical Observatory. This research made use of the XRT Data Analysis Software (XRTDAS) developed under the responsibility of the ASI Science Data Center (ASDC), Italy.
 We would also like to thank the American Association of Variable Star Observers  for supporting optical observations during the campaign.

\end{acknowledgements}

 %-------------------------Bibliography----------------------------%
 
\bibliographystyle{aa} % style aa.bst
\bibliography{./aa.bib} % your references Yourfile.bib

\begin{thebibliography}{47}
\expandafter\ifx\csname natexlab\endcsname\relax\def\natexlab#1{#1}\fi

\bibitem[{{Abdo} {et~al.}(2010){Abdo}, {Ackermann}, {Ajello}, {Atwood},
  {Baldini}, {Ballet}, {Barbiellini}, {Bastieri}, {Bechtol}, {Bellazzini}, \&
  et~al.}]{ScienceFermi}
{Abdo}, A.~A., {Ackermann}, M., {Ajello}, M., {et~al.} 2010, Science, 329, 817

\bibitem[{{Albert} {et~al.}(2008){Albert}, {Aliu}, {Anderhub}, {Antoranz},
  {Armada}, {Baixeras}, \& {Barrio}}]{RF}
{Albert}, J., {Aliu}, E., {Anderhub}, H., {et~al.} 2008, Nuclear Instruments
  and Methods in Physics Research A, 588, 424

\bibitem[{{Aleksi{\'c}} {et~al.}(2012){Aleksi{\'c}}, {Alvarez}, {Antonelli},
  {Antoranz}, {Asensio}, {Backes}, {Barrio}, {Bastieri}, {Becerra
  Gonz{\'a}lez}, {Bednarek}, {Berdyugin}, {Berger}, {Bernardini}, {Biland},
  {Blanch}, {Bock}, {Boller}, {Bonnoli}, {Borla Tridon}, {Braun}, {Bretz},
  {Ca{\~n}ellas}, {Carmona}, {Carosi}, {Colin}, {Colombo}, {Contreras},
  {Cortina}, {Cossio}, {Covino}, {Dazzi}, {de Angelis}, {de Caneva}, {de Cea
  Del Pozo}, {de Lotto}, {Delgado Mendez}, {Diago Ortega}, {Doert},
  {Dom{\'{\i}}nguez}, {Dominis Prester}, {Dorner}, {Doro}, {Elsaesser},
  {Ferenc}, {Fonseca}, {Font}, {Fruck}, {Garc{\'{\i}}a L{\'o}pez},
  {Garczarczyk}, {Garrido}, {Giavitto}, {Godinovi{\'c}}, {Hadasch},
  {H{\"a}fner}, {Herrero}, {Hildebrand}, {H{\"o}hne-M{\"o}nch}, {Hose},
  {Hrupec}, {Huber}, {Jogler}, {Kellermann}, {Klepser}, {Kr{\"a}henb{\"u}hl},
  {Krause}, {La Barbera}, {Lelas}, {Leonardo}, {Lindfors}, {Lombardi},
  {L{\'o}pez}, {L{\'o}pez-Oramas}, {Lorenz}, {Makariev}, {Maneva},
  {Mankuzhiyil}, {Mannheim}, {Maraschi}, {Mariotti}, {Mart{\'{\i}}nez},
  {Mazin}, {Meucci}, {Miranda}, {Mirzoyan}, {Miyamoto}, {Mold{\'o}n},
  {Moralejo}, {Munar-Adrover}, {Nieto}, {Nilsson}, {Orito}, {Oya}, {Paneque},
  {Paoletti}, {Pardo}, {Paredes}, {Partini}, {Pasanen}, {Pauss},
  {Perez-Torres}, {Persic}, {Peruzzo}, {Pilia}, {Pochon}, {Prada}, {Prada
  Moroni}, {Prandini}, {Puljak}, {Reichardt}, {Reinthal}, {Rhode}, {Rib{\'o}},
  {Rico}, {R{\"u}gamer}, {Saggion}, {Saito}, {Saito}, {Salvati}, {Satalecka},
  {Scalzotto}, {Scapin}, {Schultz}, {Schweizer}, {Shayduk}, {Shore},
  {Sillanp{\"a}{\"a}}, {Sitarek}, {Snidaric}, {Sobczynska}, {Spanier}, {Spiro},
  {Stamatescu}, {Stamerra}, {Steinke}, {Storz}, {Strah}, {Suri{\'c}}, {Takalo},
  {Takami}, {Tavecchio}, {Temnikov}, {Terzi{\'c}}, {Tescaro}, {Teshima},
  {Tibolla}, {Torres}, {Treves}, {Uellenbeck}, {Vankov}, {Vogler}, {Wagner},
  {Weitzel}, {Zabalza}, {Zandanel}, \& {Zanin}}]{Performance}
{Aleksi{\'c}}, J., {Alvarez}, E.~A., {Antonelli}, L.~A., {et~al.} 2012,
  Astroparticle Physics, 35, 435

\bibitem[{{Bastian} {et~al.}(1988){Bastian}, {Dulk}, \&
  {Chanmugam}}]{RelativisticElectrons}
{Bastian}, T.~S., {Dulk}, G.~A., \& {Chanmugam}, G. 1988, \apj, 324, 431

\bibitem[{{Bernl{\"o}hr} {et~al.}(2013){Bernl{\"o}hr}, {Barnacka}, {Becherini},
  {Blanch Bigas}, {Bouvier}, {Carmona}, {Colin}, {Decerprit}, {Di Pierro},
  {Dubois}, {Farnier}, {Funk}, {Hermann}, {Hinton}, {Humensky}, {Jogler},
  {Kh{\'e}lifi}, {Kihm}, {Komin}, {Lenain}, {L{\'o}pez-Coto}, {Maier}, {Mazin},
  {Medina}, {Moralejo}, {Moderski}, {Nolan}, {Ohm}, {de O{\~n}a Wilhelmi},
  {Parsons}, {Paz Arribas}, {Pedaletti}, {Pita}, {Prokoph}, {Rulten},
  {Schwanke}, {Shayduk}, {Stamatescu}, {Vallania}, {Vorobiov}, {Wischnewski},
  {Wood}, {Yoshikoshi}, {Zech}, \& {CTA Consortium}}]{PosterMC}
{Bernl{\"o}hr}, K., {Barnacka}, A., {Becherini}, Y., {et~al.} 2013, Proceedings
  of the 33rd ICRC, (arXiv:1307.2773)

\bibitem[{{Bowden} {et~al.}(1992){Bowden}, {Bradbury}, {Chadwick}, {Dickinson},
  {Dipper}, {Edwards}, {Lincoln}, {McComb}, {Orford}, {Rayner}, \&
  {Turver}}]{Narrabri92}
{Bowden}, C.~C.~G., {Bradbury}, S.~M., {Chadwick}, P.~M., {et~al.} 1992,
  Astroparticle Physics, 1, 47

\bibitem[{{Brazier} {et~al.}(1990){Brazier}, {Carraminana}, {Chadwick},
  {Dipper}, {Lincoln}, {McComb}, {Orford}, {Rayner}, {Turver}, \&
  {Williams}}]{Narrabri_Cherenkov}
{Brazier}, S.~K.~T., {Carraminana}, A., {Chadwick}, M.~P., {et~al.} 1990,
  International Cosmic Ray Conference, 4, 270

\bibitem[{{Burrows} {et~al.}(2005){Burrows}, {Hill}, {Nousek}, {Kennea},
  {Wells}, {Osborne}, {Abbey}, {Beardmore}, {Mukerjee}, {Short}, {Chincarini},
  {Campana}, {Citterio}, {Moretti}, {Pagani}, {Tagliaferri}, {Giommi},
  {Capalbi}, {Tamburelli}, {Angelini}, {Cusumano}, {Br{\"a}uninger}, {Burkert},
  \& {Hartner}}]{XRT}
{Burrows}, D.~N., {Hill}, J.~E., {Nousek}, J.~A., {et~al.} 2005, \ssr, 120, 165

\bibitem[{{Chadwick} {et~al.}(1995){Chadwick}, {Dickinson}, {Dickinson},
  {Dipper}, {Holder}, {McComb}, {Orford}, {Rayner}, {Roberts}, {Roberts},
  {Tummey}, \& {Turver}}]{Chadwick_burst}
{Chadwick}, P.~M., {Dickinson}, J.~E., {Dickinson}, M.~R., {et~al.} 1995,
  Astroparticle Physics, 4, 99

\bibitem[{{Chanmugam} \& {Brecher}(1985)}]{Acceleration}
{Chanmugam}, G. \& {Brecher}, K. 1985, \nat, 313, 767

\bibitem[{{Cheung}(2013)}]{ArxivFermi}
{Cheung}, C.~C. 2013, Fermi Symposium proceedings (arXiv:1304.3475)

\bibitem[{{Cheung} {et~al.}(2013){Cheung}, {Jean}, \& {Shore}}]{ATELCen2013}
{Cheung}, C.~C., {Jean}, P., \& {Shore}, S.~N. 2013, The Astronomer's Telegram,
  5653, 1

\bibitem[{{CTA Consortium}(2013)}]{CTA}
{CTA Consortium}. 2013, Astroparticle Physics, 43, 3

\bibitem[{{de Jager} {et~al.}(1986){de Jager}, {de Jager}, {North},
  {Raubenheimer}, {van der Walt}, \& {van Urk}}]{Nooitgedacht_Cherenkov}
{de Jager}, H.~I., {de Jager}, O.~C., {North}, A.~R., {et~al.} 1986, South
  African Journal of Physics, 9, 107

\bibitem[{{de Jager}(1994)}]{deJager_pulsed_ULs}
{de Jager}, O.~C. 1994, \apj, 436, 239

\bibitem[{{de Jager} {et~al.}(1994){de Jager}, {Meintjes}, {O'Donoghue}, \&
  {Robinson}}]{deJager94}
{de Jager}, O.~C., {Meintjes}, P.~J., {O'Donoghue}, D., \& {Robinson}, E.~L.
  1994, \mnras, 267, 577

\bibitem[{{de Jager} {et~al.}(1989){de Jager}, {Raubenheimer}, \&
  {Swanepoel}}]{H-test}
{de Jager}, O.~C., {Raubenheimer}, B.~C., \& {Swanepoel}, J.~W.~H. 1989, \aap,
  221, 180

\bibitem[{{Fomin} {et~al.}(1994){Fomin}, {Stepanian}, {Lamb}, {Lewis}, {Punch},
  \& {Weekes}}]{Wobble}
{Fomin}, V.~P., {Stepanian}, A.~A., {Lamb}, R.~C., {et~al.} 1994, Astroparticle
  Physics, 2, 137

\bibitem[{{Friedjung}(1997)}]{Distance}
{Friedjung}, M. 1997, \na, 2, 319

\bibitem[{{Gehrels} {et~al.}(2004){Gehrels}, {Chincarini}, {Giommi}, {Mason},
  {Nousek}, {Wells}, {White}, {Barthelmy}, {Burrows}, {Cominsky}, {Hurley},
  {Marshall}, {M{\'e}sz{\'a}ros}, {Roming}, {Angelini}, {Barbier}, {Belloni},
  {Campana}, {Caraveo}, {Chester}, {Citterio}, {Cline}, {Cropper}, {Cummings},
  {Dean}, {Feigelson}, {Fenimore}, {Frail}, {Fruchter}, {Garmire}, {Gendreau},
  {Ghisellini}, {Greiner}, {Hill}, {Hunsberger}, {Krimm}, {Kulkarni}, {Kumar},
  {Lebrun}, {Lloyd-Ronning}, {Markwardt}, {Mattson}, {Mushotzky}, {Norris},
  {Osborne}, {Paczynski}, {Palmer}, {Park}, {Parsons}, {Paul}, {Rees},
  {Reynolds}, {Rhoads}, {Sasseen}, {Schaefer}, {Short}, {Smale}, {Smith},
  {Stella}, {Tagliaferri}, {Takahashi}, {Tashiro}, {Townsley}, {Tueller},
  {Turner}, {Vietri}, {Voges}, {Ward}, {Willingale}, {Zerbi}, \&
  {Zhang}}]{Swift}
{Gehrels}, N., {Chincarini}, G., {Giommi}, P., {et~al.} 2004, \apj, 611, 1005

\bibitem[{{Hays} {et~al.}(2013){Hays}, {Cheung}, \& {Ciprini}}]{ATEL}
{Hays}, E., {Cheung}, T., \& {Ciprini}, S. 2013, The Astronomer's Telegram,
  5302

\bibitem[{{Hobbs} {et~al.}(2006){Hobbs}, {Edwards}, \& {Manchester}}]{TEMPO2}
{Hobbs}, G.~B., {Edwards}, R.~T., \& {Manchester}, R.~N. 2006, \mnras, 369, 655

\bibitem[{{Ikhsanov}(1998)}]{Ikhsanov_pulsar}
{Ikhsanov}, N.~R. 1998, \aap, 338, 521

\bibitem[{{Ikhsanov} \& {Biermann}(2006)}]{EWD-model}
{Ikhsanov}, N.~R. \& {Biermann}, P.~L. 2006, \aap, 445, 305

\bibitem[{{Kitaguchi} {et~al.}(2014){Kitaguchi}, {An}, {Beloborodov},
  {Gotthelf}, {Hayashi}, {Kaspi}, {Rana}, {Boggs}, {Christensen}, {Craig},
  {Hailey}, {Harrison}, {Stern}, \& {Zhang}}]{Nustar}
{Kitaguchi}, T., {An}, H., {Beloborodov}, A.~M., {et~al.} 2014, \apj, 782, 3

\bibitem[{{Kuijpers} {et~al.}(1997){Kuijpers}, {Fletcher}, {Abada-Simon},
  {Horne}, {Raadu}, {Ramsay}, \& {Steeghs}}]{Kuijpers}
{Kuijpers}, J., {Fletcher}, L., {Abada-Simon}, M., {et~al.} 1997, \aap, 322,
  242

\bibitem[{{Lang} {et~al.}(1998){Lang}, {Buckley}, {Carter-Lewis}, {Catanese},
  {Cawley}, {Colombo}, {Connaughton}, {Fegan}, {Finley}, {Gaidos},
  {Gillanders}, {Hillas}, {Kertzman}, {Krennrich}, {Lessard}, {Moriarty},
  {Quinn}, {Rose}, {Sembroski}, \& {Weekes}}]{AEAqr_Whipple}
{Lang}, M.~J., {Buckley}, J.~H., {Carter-Lewis}, D.~A., {et~al.} 1998,
  Astroparticle Physics, 9, 203

\bibitem[{Mardia(1972)}]{Rayleigh_test}
Mardia, K.~V. 1972, Statistics of Directional Data. Academic Press, New York.

\bibitem[{{Mauche}(2006)}]{MaucheBreak}
{Mauche}, C.~W. 2006, \mnras, 369, 1983

\bibitem[{{Mauche}(2009)}]{Mauche_2009}
{Mauche}, C.~W. 2009, \apj, 706, 130

\bibitem[{{Mauche} {et~al.}(2012){Mauche}, {Abada-Simon}, {Desmurs}, {Dulude},
  {Ioannou}, {Neill}, {Price}, {Sidro}, {Welsh}, \& {AAVSO CBA}}]{MaucheMWL}
{Mauche}, C.~W., {Abada-Simon}, M., {Desmurs}, J.-F., {et~al.} 2012, \memsai,
  83, 651

\bibitem[{{Meintjes} \& {de Jager}(2000)}]{PropellerModel}
{Meintjes}, P.~J. \& {de Jager}, O.~C. 2000, \mnras, 311, 611

\bibitem[{{Meintjes} {et~al.}(1994){Meintjes}, {de Jager}, {Raubenheimer},
  {Nel}, {North}, {Buckley}, \& {Koen}}]{Meintjes94}
{Meintjes}, P.~J., {de Jager}, O.~C., {Raubenheimer}, B.~C., {et~al.} 1994,
  \apj, 434, 292

\bibitem[{{Meintjes} {et~al.}(2012){Meintjes}, {Oruru}, \&
  {Odendaal}}]{Meintjes2012}
{Meintjes}, P.~J., {Oruru}, B., \& {Odendaal}, A. 2012, \memsai, 83, 643

\bibitem[{{Nilsson}(2014)}]{KVA_software}
{Nilsson}, K. 2014, in preparation

\bibitem[{{Patterson}(1979)}]{Patterson79}
{Patterson}, J. 1979, \apj, 234, 978

\bibitem[{{Patterson}(1994)}]{Patterson94}
{Patterson}, J. 1994, \pasp, 106, 209

\bibitem[{{Patterson} {et~al.}(1980){Patterson}, {Branch}, {Chincarini}, \&
  {Robinson}}]{X-ray_pulsation}
{Patterson}, J., {Branch}, D., {Chincarini}, G., \& {Robinson}, E.~L. 1980,
  \apjl, 240, L133

\bibitem[{{Rolke} {et~al.}(2005){Rolke}, {L{\'o}pez}, \& {Conrad}}]{Rolke}
{Rolke}, W.~A., {L{\'o}pez}, A.~M., \& {Conrad}, J. 2005, Nuclear Instruments
  and Methods in Physics Research A, 551, 493

\bibitem[{{Sidro} {et~al.}(2008){Sidro}, {Cortina}, {Mauche}, \& {et
  al.}}]{Sidro}
{Sidro}, N., {Cortina}, J., {Mauche}, C.~W., \& {et al.} 2008, in International
  Cosmic Ray Conference, Vol.~2, International Cosmic Ray Conference, 715--718

\bibitem[{{Takalo} {et~al.}(2008){Takalo}, {Nilsson}, {Lindfors},
  {Sillanp{\"a}{\"a}}, {Berdyugin}, \& {Pasanen}}]{KVA}
{Takalo}, L.~O., {Nilsson}, K., {Lindfors}, E., {et~al.} 2008, in American
  Institute of Physics Conference Series, ed. F.~A. {Aharonian}, W.~{Hofmann},
  \& F.~{Rieger}, Vol. 1085, 705--707

\bibitem[{{Terada}(2013)}]{Cosmic_electrons}
{Terada}, Y. 2013, Thirteenth Marcel Grossmann Meeting (arXiv:1306.4053)

\bibitem[{{Terada} {et~al.}(2008){Terada}, {Hayashi}, {Ishida}, {Mukai},
  {Dotani}, {Okada}, {Nakamura}, {Naik}, {Bamba}, \&
  {Makishima}}]{SuzakuTerada}
{Terada}, Y., {Hayashi}, T., {Ishida}, M., {et~al.} 2008, \pasj, 60, 387

\bibitem[{{Warner}(2003)}]{CVs}
{Warner}, B. 2003, {Cataclysmic Variable Stars} (Cambridge University Press)

\bibitem[{{Welsh} {et~al.}(1998){Welsh}, {Horne}, \& {Gomer}}]{Doppler}
{Welsh}, W.~F., {Horne}, K., \& {Gomer}, R. 1998, \mnras, 298, 285

\bibitem[{{Wynn} {et~al.}(1997){Wynn}, {King}, \& {Horne}}]{Propeller}
{Wynn}, G.~A., {King}, A.~R., \& {Horne}, K. 1997, \mnras, 286, 436

\bibitem[{{Zanin} {et~al.}(2013){Zanin}, {Carmona}, {Sitarek}, {Colin},
  {Frantzen}, {Gaug}, {Lombardi}, \& {for the MAGIC collaboration}}]{MARS}
{Zanin}, R., {Carmona}, E., {Sitarek}, J., {et~al.} 2013, Proceedings of the
  ICRC 2013, id 773

\end{thebibliography}

\end{document}